\documentclass[twocolumn,showpacs,preprintnumbers,amsmath,amssymb,pra]{revtex4}
\usepackage{epsf,graphicx,dcolumn,bm}
\newcommand{\bq}{\begin{equation}}
\newcommand{\eq}{\end{equation}}
\newcommand{\ba}{\begin{array}}
\newcommand{\ea}{\end{array}}
\newcommand{\bea}{\begin{eqnarray}}
\newcommand{\eea}{\end{eqnarray}}
\newcommand {\tab} {\hspace*{2em}}
\newcommand{\fa}{\forall}

\begin{document}
\title{$N$-Qubit $W$ States are Determined by their Bipartite
Marginals}
\author{Preeti Parashar}
\email{parashar@isical.ac.in}
\author{Swapan Rana}
\email{swapan_r@isical.ac.in} \affiliation{Physics and Applied
Mathematics Unit, Indian Statistical Institute, 203 B.T. Road,
Kolkata-700108}
\date{\today}

\begin{abstract}
We prove that the most general $W$ class of $N$-qubit states are
uniquely determined among arbitrary states (pure or mixed) by just
their bipartite reduced density matrices. Moreover, if we consider
only pure states, then $(N-1)$ of them are shown to be sufficient.
\end{abstract}
 \maketitle

 Understanding the structure of multi party quantum correlations
is an important issue in quantum information. Different types of
correlations that a multi-partite state exhibits lead to
classification and characterization of quantum states. Essentially the
study reveals peculiarity of quantum correlation as compared to
its classical analogue. For example, contrary to the quantum case,
two different generic trivariate classical probability distributions
can have the same bivariate marginals.

 One of the basic questions concerning the study of quantum correlations
is whether higher order correlations follow from lower order ones.
This was first addressed by Linden, Popescu and Wootters
\cite{LPW} where they proved that a generic $3$-qubit pure state
is uniquely determined by its two-party reduced states. Extending
this proof to more parties having any finite dimension, Linden and
Wootters \cite{LW} have given some bounds on the number of reduced
parties. An alternative technique to prove the result of
\cite{LPW} was proposed by Diosi \cite{Diosi} by making use of the
Schmidt decomposition. It was also shown in \cite{LPW} that the
only exceptional class of $3$-qubit pure states not determined
uniquely by its bipartite marginals is the Greenberger-
Horne-Zeilinger (GHZ) class $(a|000\rangle+b|111\rangle)$. This
implies that only these states carry information at the three
qubit level, since their correlations are not reducible.
Generalizing this result, Walck and Lyons \cite{WL} have shown
that GHZ is the only class of $N$-qubit states which are not
determined by their $(N-1)$-partite marginals. But generically the
``Parts'' can determine the ``Whole''. Recently, a quantitative
measure of the degree of irreducible $K$-particle correlations in
an $N$-particle state, based on the maximal entropy construction
has been defined by Zhou \cite{Zhou}, in particular for stabilizer
and generalized GHZ states.

 It is known \cite{Dur} that under Stochastic Local Operation and
Classical Communication (SLOCC), there exists two inequivalent
classes of $3$-qubit genuinely entangled pure states - the GHZ
class and the $W$ class $( a|001\rangle + b|010\rangle +
c|100\rangle )$. These later states have an interesting property
that their entanglement exhibits maximum robustness against the
loss of one qubit. This means that the bipartite entanglement left
in the system can still be used as a resource to perform
information-theoretic tasks, even in the absence of cooperation
from the third party. Moreover, the third party cannot destroy the
residual entanglement, thereby making $W$ state especially useful
for secure communication \cite{ASD}. Also the reduced bipartite
entangled state can be brought arbitrarily close to a Bell state
by means of a filtering measurement \cite{Gisin}. Motivated by
these special features, we investigate the most general $W$ class
of states in terms of their reducible correlations.

 It would be interesting to find classes of
states which could be determined by fewer than $(N-1)$-partite
marginals. However, this task becomes quite challenging since the
known techniques cannot be applied to such situations. The present
Communication is a first attempt in this direction. We prove that
the $N$-qubit pure $W$ class of states
($|W\rangle_N=a_1|0...1\rangle + ... +a_N|1...0\rangle$, ~~all
$a_i$ complex) are uniquely determined by their \textit
{two-party} reduced density matrices. There does not exist any
other pure or mixed $N$-qubit state sharing the same bipartite
marginals.

 Before proceeding to the proof, we will adopt some notations, for
easier understanding.

Since a Hermitian matrix $A$ is usually identified by the elements
$a_{ij}, \forall i\leq j $, a density matrix (written in some
basis) is necessarily identified by its upper-half elements
($a_{ij}, \forall i< j $) together with the diagonal elements
$a_{ii}$. The lower-half entries ($a_{ij}, \forall i> j $) are
redundant as they are just complex conjugates of the upper ones.
Therefore, we can write a general (possibly mixed) $N$-qubit density matrix in
standard Computational Basis (CB) as \cite{rmat}
\bq
\rho^{12...N}_M~=~\sum_{i=0}^{2^N-1}\sum_{j=i}^{2^N-1}r_{ij}|B_N(i)\rangle\langle
B_N(j)|\eq where $B_N(i)$ is the binary equivalent of the decimal
number $i$ in an $N$-bit string.

Another key observation is that to compute the reduced density
matrix (RDM) (of some parties) from an $N$-qubit pure state (in
CB)\bq|\psi\rangle_N = \sum_{i=0}^{2^N-1}a_i|B_N(i)\rangle,~\eq we
need to find the \emph{expressions} for only the diagonal entries
of that RDM in terms of the state coefficients. Note that we
require the expressions (i.e., the complex numbers appearing in
the sum) and not the final calculated values (which are always
real) of the diagonals . All other non-diagonal entries will be
determined from these expressions. To see it explicitly, let us
consider the m-partite marginal $\rho^{i_1i_2...i_m}$ of $\rho =
|\psi\rangle_N\langle\psi|$, where $\rho^{i_1i_2...i_m} =
Tr_{i_{m+1}i_{m+2}...i_N}(\rho)$. This RDM can be written as \bq
\rho^{i_1i_2...i_m}~=~\sum_{i=0}^{2^m-1}\sum_{j=i}^{2^m-1}r_{ij}|B_m(i)\rangle\langle
B_m(j)|.\eq Clearly it has $2^m$ diagonal entries ($r_{ii}, \fa
i=0(1)2^m-1)$. To find the expression of $r_{ii}$, we should first
write the decimal number $i$ by putting $t_1$, $t_2$,....,$t_m$
(each $t_j$ can take value 0/1) in an $m$-bit binary string. Then
taking an $N$-bit string we put the $t_j$'s at $i_j$th places
respectively (see Fig.1) and
\begin{figure}[ht]
    \begin{center}
            \resizebox{6.5cm}{1.5cm}{\includegraphics*[4cm,12.3cm][20cm,15cm]{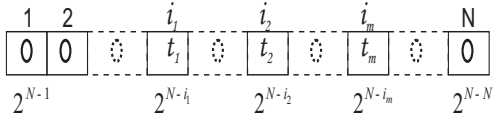}}
            \caption{\footnotesize Least-suffix $=\sum_{j=1}^mt_j2^{N-i_j}$}\label{fig1}
     \end{center}
\end{figure}
then fill the remaining ($N-m$) places of the string arbitrarily
by 0 and 1. These ($N-m$) places can be filled in $2^{N-m}$
different ways. Hence each $r_{ii}$ will be the sum of $2^{N-m}$
number of terms (each term is of the form $|a_k|^2$, where the
decimal number $k$ in the suffix is obtained by converting the
binary arrangement to the decimal number). If we write these terms
in increasing order of the suffixes, the first term in the
expression of $r_{ii}$ will be $|a_{\sum_{j=1}^mt_j2^{N-i_j}}|^2$.
(This least suffix is obtained by converting the binary number
formed by putting $t_j$'s at $i_j$th places and filling all other
places of the $N$-bit string by 0). Once we know the expression
for $r_{ii}$, the expression for $r_{ij}$ follows trivially. To be
explicit, say $r_{ii} = \sum_{k=0}^{2^{N-m}-1}|p_k|^2$ and $r_{jj}
= \sum_{k=0}^{2^{N-m}-1}|q_k|^2$ ($p_k$, $q_k$ are some $a_l$'s
with the suffix $l$ in increasing order), then $r_{ij} =
\sum_{k=0}^{2^{N-m}-1}p_k\bar{q_k}$. Note that $r_{ii}$, $r_{jj}$
and $r_{ij}$ are the co-efficient of $|B_m(i)\rangle\langle
B_m(i)|$, $|B_m(j)\rangle\langle B_m(j)|$  and
$|B_m(i)\rangle\langle B_m(j)|$ respectively in the RHS of (3).
(See Appendix).

 Now we will prove our claim:
\textbf{\textit{\boldmath{N}-Qubit Pure \boldmath{W} States are
Determined by their Bipartite Marginals}}.

To prove this we will show that there exists no $N$-qubit state
(rather Density Matrix) having the same bipartite marginals as
those of an $N$-qubit $W$-state \bq |W\rangle_N =
\sum_{i=0}^{N-1}w_{2^i}|B_N(2^i)\rangle \eq with
$\sum_{i=0}^{N-1}|w_{2^i}|^2 = 1$ other than \bq
|W\rangle_N\langle W|~=~\sum_{i=0}^{N-1}\sum_{j=i}^{N-1}w_{2^i}
\bar w_{2^j}|B_N(2^i)\rangle\langle B_N(2^j)|.\eq

\textbf{\textit{Proof}}

1. \tab Let us first find the bipartite marginals $\rho^{JK}_W$ of
$|W\rangle_N$. As discussed above, we need to calculate the
\emph{expressions} for only the 4 diagonal elements $r_{ii}, i =
0(1)3$. Since the CB states in (4) contain just one 1, $r_{33} =
0$ in all $\rho^{JK}_W$. To find $r_{11}$ (which is the
co-efficient of $|01\rangle\langle01|$), we should put one 1 at
$K$th place of N-bit string and there is exactly one such CB in
(4) having co-efficient $w_{2^{N-K}}$. Hence $r_{11} =
|w_{2^{N-K}}|^2$. Similarly, $r_{22} = |w_{2^{N-J}}|^2$. $r_{00}$
can be obtained from the normalization condition. Thus  \bea
\rho^{JK}_W &=&~ (1-|w_{2^{N-J}}|^2-|w_{2^{N-K}}|^2)|00\rangle
\langle00| \nonumber
\\&~&+
|w_{2^{N-K}}|^2|01\rangle \langle01|+|w_{2^{N-J}}|^2|10\rangle
\langle10|\nonumber
\\&~&+w_{2^{N-K}}\bar w_{2^{N-J}}|01\rangle \langle10| \eea
 Written in
matrix form,  \bq \rho^{JK}_W=\left[%
\begin{array}{cccc}
r_{00} & 0 & 0 & 0 \\
   & |w_{2^{N-K}}|^2 & w_{2^{N-K}}\bar w_{2^{N-J}} & 0 \\
   &  & |w_{2^{N-J}}|^2 & 0 \\
   & &  & 0 \\
\end{array}%
\right]\eq where $r_{00}= 1-|w_{2^{N-J}}|^2-|w_{2^{N-K}}|^2$.

2. \tab Now let us suppose that there exists an $N$-qubit density
matrix $\rho^{12...N}_M$ given by (1), which
has the same bipartite marginals as $|W\rangle_N$. In matrix form,
all bipartite marginals $\rho^{JK}_M$ of $\rho^{12...N}_M$ will
have four diagonal elements $d_i^{JK}$ at ($i$,$i$) positions, $i
= 1(1)4$. Let us first calculate the diagonal element $d_4^{JK}$
at (4,4) position of $\rho^{JK}_M$. Obviously, this will be a sum
of $2^{N-2}$ number of diagonal terms ($r_{ii}$) of
$\rho^{12...N}_M$. To see explicitly which $r_{ii}$'s will appear
in the sum, we observe that the suffixes $i$ will vary over the
decimal numbers obtained by
\begin{figure}[ht]
     \begin{center}
         \resizebox{6.5cm}{1.5cm}{\includegraphics*[4cm,12.3cm][18cm,15cm]{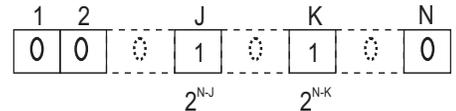}}
         \caption{\footnotesize Least-suffix in $d^{JK}_4$ is $2^{N-J}+2^{N-K}$}\label{fig2}
     \end{center}
\end{figure}
\noindent converting the N-bit binary numbers having 1 fixed
at $J$th \& $K$th places and arbitrarily 0/1 at the remaining
(N-2) places. (As an illustration, Fig.2 shows how to get the least suffix there).
Hence the terms $r_{ii}$'s for the suffixes $i = 0$
and  $i = 2^{j-1}~ \fa j = 1(1)N$ will not appear in the
expression of $d_4^{JK}$ for any $J,K$ as the binary
representation of these numbers can have at most one 1 (but we
need at least two 1).

3a.\tab Now comparing the (4,4) position of $\rho^{JK}_M$ and
$\rho^{JK}_W$, it follows that $d_4^{JK}$ = 0 $\fa~J,K$. Since the
diagonal elements $r_{ii}\geq 0 ~\fa i$ \cite{PSD}, it follows
that all $r_{ii}$ appearing in the expression of $d_4^{JK}$ should
individually be zero. Hence it implies (from step 2) that the only
non-zero diagonal elements of $\rho^{12...N}_M$ are $r_{ii}$ for
$i=0~  \& ~ i=2^{j-1}~\fa j=1(1)N$.

3b.\tab Next considering $d^{1K}_3$ for any fixed $K$ and
comparing with the element at (3,3) position of $\rho^{1K}_W$, we
get $r_{2^{N-1}2^{N-1}}=|w_{2^{N-1}}|^2$. Similarly comparing
(2,2) position of $\rho^{1K}_M$ and $\rho^{1K}_W$, it follows that
$r_{2^{N-K}2^{N-K}}=|w_{2^{N-K}}|^2~\fa K=2(1)N$ i.e.,
$r_{2^i2^i}=|w_{2^i}|^2~\fa i=0$ to ($N-2$).

3c.\tab Finally, from normalization condition
$\sum_{i=0}^{2^N-1}r_{ii}=1=\sum_{i=0}^{N-1}|w_{2^i}|^2$, we get
$r_{00}=0$. Thus collecting the results of steps 3a \& 3b we get
\bq r_{2^i2^i} = |w_{2^i}|^2, \fa i=0(1)N-1\eq and all other
$r_{ii}$ in $\rho^{12...N}_M$ are zero.

4.\tab Now we will use the fact that if a diagonal element of a
positive semidefinite (PSD) matrix is zero, then all elements in
the row and column containing that element should be zero
\cite{book}. Hence (using the final result of step 3c) it follows
that \bq
\rho^{12...N}_M~=~\sum_{i=0}^{N-1}\sum_{j=i}^{N-1}r_{2^i2^j}|B_N(2^i)\rangle\langle
B_N(2^j)|\eq where $r_{2^i2^i}$ are given by (8). Note that
$\rho^{12...N}_M$ in eqn.(9) reduces to the same form as
$|W\rangle_N\langle W|$ given in (5).

5. \tab The non-diagonal element $r_{2^i2^j}$ at ($2^i,2^j$) place
of $\rho^{12...N}_M$ is found to be $w_{2^i} \bar w_{2^j}$ by
comparing the elements at (2,3) position of $\rho^{(N-j)(N-i)}_M$
and $\rho^{(N-j)(N-i)}_W$. Thus $r_{2^i2^j}=w_{2^i} \bar w_{2^j}$
and hence it follows that $\rho^{12...N}_M~=~|W\rangle_N\langle
W|$.
\hfill$\blacksquare$\\

In the above analysis considering mixed states, we needed all
$^NC_2$ number of bipartite marginals to determine $|W\rangle_N$.
However, if restricted  to pure states only, then it can be shown
that we require only ($N-1$) bipartite marginals $\rho^{JK}_W$,
where one of the parties is fixed and the other varies over
all the remaining ($N-1$) parties. Without loss of generality, we
can take the first party as the fixed one and thus
$\rho^{1K}_W~\fa~K=2(1)N$ are sufficient to determine
$|W\rangle_N$.

We will now prove this second claim: \textbf{\textit{Among pure
states, \boldmath{N}-Qubit \boldmath{W} States are Determined by
their \boldmath{N-1} Bipartite Marginals
\boldmath{$\rho^{1K}_W,~K=2(1)N$}}}.

To prove this claim it suffices to show that there exists no
$N$-qubit pure state $|\psi\rangle_N$ given by eqn.(2) having the
same bipartite marginals.

\textbf{\textit{Proof}}

 1.\tab First compare the entries at (4,4) position of
$\rho^{1K}_\psi$ and $\rho^{1K}_W~\fa~K=2(1)N$. This will make
the last (arranged in increasing order of the suffix) $2^{N-1}-1$
number of coefficients vanish, i.e., $a_i=0~\fa i=2^{N-1}+1~
\mbox{to}~ 2^N-1$ [The decimal numbers generated by placing 1
at 1st \& Kth places and filling all others places arbitrarily by
0/1 $\fa K=2(1)N$].

2.\tab Now look at the entries at (3,3) position of
 $\rho^{1K}_\psi$. For all values of $K$ there is only one term $|a_{2^{N-1}}|^2$ in
 common and this is the term  with least suffix. (The common bit 1 at Most Significant
 Place has decimal value $2^{N-1}$). All remaining $2^{N-2}-1$ terms of each
 $\rho^{1K}_\psi$are nothing but the terms considered in step 1
 (as the suffixes generated will be $>$ $2^{N-1}$). So the only
 non-zero term at (3,3) position of each $\rho^{1K}_\psi$ is the
 common term $|a_{2^{N-1}}|^2$ which when compared with
 $\rho^{1K}_W$ gives\bq |a_{2^{N-1}}|=|w_{2^{N-1}}|\eq

3.\tab The element at (2,3) place of $\rho^{1K}_\psi$
is the sum of products of corresponding $a_i$'s
appearing at (2,2) position and complex conjugate of $a_k$'s
appearing at (3,3) position. The term with least suffix at (2,2)
position is $|a_{2^{N-K}}|^2$. Hence (by step 2, $|a_{2^{N-1}}|^2$
is the only non-zero term at (3,3) position and this is the least
suffix term there) comparing with $\rho^{1K}_W$, \begin{eqnarray}
a_{2^{N-K}}\bar a_{2^{N-1}}&=& w_{2^{N-K}}\bar
w_{2^{N-1}}\nonumber
\\\Rightarrow~~~~~
\frac{a_{2^{N-K}}}{w_{2^{N-K}}}&=&\frac{\bar
w_{2^{N-1}}}{\bar a_{2^{N-1}}}~=~e^{i\phi}\nonumber\\
\Rightarrow~~~~~~a_{2^{N-K}}&=&e^{i\phi}w_{2^{N-K}},~\fa K=1(1)N
\end{eqnarray} where
$\phi=arg(a_{2^{N-1}})-arg(w_{2^{N-1}})$ is a fixed number.

4.\tab Next consider the entries at (2,2) position of
$\rho^{1K}_\psi$. For all $K$, the sum there starts with the least
suffix term $|a_{2^{N-K}}|^2=|w_{2^{N-K}}|^2$ (by (11)). So
comparison with $\rho^{1K}_W$ will make the remaining
$2^{N-2}-1$ terms vanish.

5.\tab Finally, the normalization condition gives $a_0=0$.
Therefore, \begin{eqnarray}|\psi\rangle_N &=&
\sum_{i=0}^{2^N-1}a_i|B_N(i)\rangle\nonumber \\
&=&e^{i\phi}\sum_{i=1}^{N}w_{2^i}|B_N(2^i)\rangle \nonumber \\
&=&|W\rangle_N\nonumber\end{eqnarray} \hfill$\blacksquare$

In conclusion, we have shown that the $N$-qubit $W$ class of
states are uniquely determined by just their bipartite marginals.
This reveals that we do not require information regarding the
reduced states beyond the bipartite level. In other words, the
$N$-party correlations in $W$ states are reducible to 2-party
correlations. Therefore, any higher party $W$ state can be
constructed from bipartite reduced density matrices. Recently, a
lot of effort has been devoted to experimental generation of
multi-qubit $W$ states by different methods \cite{rec}. We hope
that our result may be useful to make the process easier as it is
sufficient to consider only bipartite marginals. In a slightly
different context, such investigations are being carried out in
molecular physics, viz., construction of density matrices of an
$N$-electron system from its bipartite marginals (see \cite{LW}
for references).

 Another notable point is that for $|W\rangle_N$, the loss of any
$N-2$ parties still leaves it in a
bipartite entangled state\\
$\rho^{JK}_W=|\psi^+\rangle_{JK} \langle
\psi^+|+(1-|w_{2^{N-J}}|^2-|w_{2^{N-K}}|^2)
|00\rangle \langle 00|$\\
where $|\psi^+\rangle_{JK}= w_{2^{N-K}} |01\rangle + w_{2^{N-J}}
|10\rangle $, which can be distilled. It has been suggested in the
literature \cite{Dur} that the entanglement of $W$ state is
readily bipartite. Our result confirms this. As $|W\rangle_N$ can
be determined uniquely from bipartite marginals,  any property of
the whole state should be characterized by these marginals. Thus
the total entanglement in $W$ state should essentially be
characterized by the bipartite entanglement present in it. Also it
is very likely that there exists a close relation between the
determination of $|W\rangle_N$ from bipartite marginals and the
saturation of the general monogamy inequality \cite{CKW, OV} for
bipartite entanglement. However, further investigation is required
in order to establish this.

 It is natural to ask whether $|W\rangle_N$ can uniquely
be determined by \textit{arbitrary} $(N-1)$ bipartite marginals.
We have verified that the answer is in the affirmative only for
$N=3, 4$. This is due to the fact that up to $N=4$, any set of
$(N-1)$ number of bipartite marginals covering the parties would
automatically correlate all the parties. However, for $N\ge 5$,
there always exist some set which would not correlate all the
parties. For example, for $N=5$, the set $\{\rho^{12}, \rho^{13},
\rho^{23}, \rho^{34} \}$ cannot determine $|W\rangle_5$ uniquely.

 Another interesting issue would be to find if $(N-1)$ is the
optimal number or $|W\rangle_N$ can be determined by less number
of bipartite marginals. It is tempting to think that it is
sufficient to take $[\frac{N+1}{2}]$ bipartite marginals (since
this is the minimum number required to cover all parties).
However, this is not the case. For even $N$, one particular
coverage can be achieved by taking the marginals $\rho^{J(J+1)}
~~, J=1(2)(N-1)$. But such a coverage does not determine
$|W\rangle_N$ uniquely. For example \bea |W\rangle _4=
a|0001\rangle+b| 0010\rangle+c| 0100\rangle+d|1000\rangle~\mbox{and}\tab \nonumber\\
|W^{\prime}\rangle_4=e^{i\theta}(a|0001\rangle
  +b|0010\rangle)+e^{i\phi}(c|0100\rangle +d|1000\rangle) \nonumber
  \eea share the same $\rho^{12}$ and $\rho^{34}$. The reason for this indeterminacy
is that $\rho^{12}$ and $\rho^{34}$ can be viewed as local marginals and they do not
capture the entire correlation. So it is necessary to take into account one more marginal
to correlate them. Similar argument holds for the  general case.

 We have presented the first example of a class of $N$-party quantum states
determinable from fewer than $(N-1)$-partite marginals. It is also
worth finding other classes determinable from $K$-partite reduced
matrices for $K < (N-1)$. The method developed here can be readily
applied to identify states which exhibit $K$-party correlations.
It also needs to be checked whether $W$ is the only class whose
correlations can be reduced to $2$-party level.

 We hope this work would provide some insight into the
general problem of reducibility of multi-partite quantum
correlations, and shed light on its close connection with quantum
entanglement.

We thank Sandu Popescu for bringing to our attention the paper
\cite{LPW} which has inspired the present work.

\hrulefill

\leftline{\bf Appendix} Here we will explicitly show how one can
obtain the off diagonals from the \emph{expression} of the
diagonal elements, by considering a simple example.

Let us determine the bipartite RDM $\rho^{12}_{\psi}$ from\\
$|\psi\rangle=\frac{1}{\sqrt{2}}(|001\rangle+i|111\rangle)$.

In matrix form,
$\rho=|\psi\rangle\langle\psi|=[r_{ij}]^{i,j=0,\ldots,7}_{i\leq
j}$ and $\rho^{12}_{\psi}=[R_{ij}]^{i,j=0,\ldots,3}_{i\leq j}$.
Then the diagonal terms are given by\\
$R_{00}=r_{00}+r_{11}=|0|^2+|\frac{1}{\sqrt{2}}|^2=\frac{1}{2}$\\
$R_{11}=r_{22}+r_{33}=|0|^2+|0|^2=0$\\
$R_{22}=r_{44}+r_{55}=|0|^2+|0|^2=0$\\
$R_{33}=r_{66}+r_{77}=|0|^2+|\frac{i}{\sqrt{2}}|^2=\frac{1}{2}$

Now from the explicit expression (and not from their
\emph{calculated final real values} $\frac{1}{2}$ and $0$) of
these diagonal entries, we can get the off-diagonal terms. For
example, $R_{23}=\sum_{\mbox {corresponding terms~}}$ (complex
number appearing in $R_{22}$)(conjugate of complex number
appearing in $R_{33})=0.(\bar{0})+0.(\bar{\frac{i}{\sqrt{2}}})=0$\\
Similarly,
$R_{03}=0.(\bar{0})+\frac{1}{\sqrt{2}}.(\bar{\frac{i}{\sqrt{2}}})=-\frac{i}{2}$.


\begin{thebibliography}{99}
\bibitem{LPW} N. Linden, S. Popescu and W. K. Wootters, Phys. Rev. Lett. {\bf 89}, 207901 (2002).

\bibitem{LW} N. Linden and W. K. Wootters, Phys. Rev. Lett. {\bf 89}, 277906 (2002).

\bibitem{Diosi}L. Diosi, Phys. Rev. A {\bf 70}, 010302(R) (2004).

\bibitem{WL} S. N. Walck and D. W. Lyons, Phys. Rev. Lett. {\bf 100},
050501 (2008); S. N. Walck and D. W. Lyons, Phys. Rev. A {\bf 79},
032326 (2009).

\bibitem{Zhou} D.L. Zhou, Phys. Rev. Lett. {\bf 101}, 180505 (2008).

\bibitem{Dur} W. Dur, G. Vidal and J.I. Cirac, Phys. Rev. A {\bf
62}, 062314 (2000).

\bibitem{ASD} A. Sen(De), \textit{et al.} Phys. Rev. A {\bf 68}, 062306
(2003).
\bibitem{Gisin} N. Gisin, Phys. Lett. A, {\bf 210}, 151 (1996).

\bibitem{rmat} To avoid technical difficulties, one may think of the\\following
matrix form: $\left[%
\begin{array}{cccc}
  r_{00} & r_{01} & \ldots  & r_{02^N-1} \\
   & r_{11} & \ldots & r_{12^N-1} \\
   & & \ddots & \ldots\\
   &   & & r_{2^N-12^N-1} \\
\end{array}%
\right]$

\bibitem{PSD} If possible let $d_i<0$ in a PSD matrix $A$. Then taking
$|\psi\rangle={[0,\ldots,1,\ldots,0]}^T,~\langle\psi|A|\psi\rangle=d_i<0$,
a contradiction.

\bibitem{book} \emph{Handbook of Linear Algebra}, edited by L. Hogben,
 (Chapman and Hall, London/CRC, Cleveland, 2007), chap.8, p. 7.

\bibitem{rec}  G-C Guo  and Y-S Zhang, Phys. Rev. A {\bf 65}, 054302 (2002);~
           X. Wang,  M. Feng and B.C. Sanders, \emph{ibid.} {\bf 67}, 022302 (2003);~
           H. Mikami, Y. Li and T. Kobayashi, \emph{ibid.} {\bf 70}, 052308 (2004);~
           M. Eibl \textit{et al.} Phys. Rev. Lett. {\bf 92}, 077901 (2004);~
           H. Mikami \textit{et al.}  \emph{ibid.} {\bf 95}, 150404 (2005).

\bibitem{CKW} V. Coffman, J. Kundu and W.K. Wootters, Phys. Rev. A
{\bf 61}, 052306 (2000).

\bibitem{OV} T.J. Osborne and F. Verstraete, Phys. Rev. Lett. {\bf
96}, 220503 (2006).

\end{thebibliography}
\end{document}